\documentclass[manuscript]{aastex61}

\usepackage{url}
\usepackage{tabularx}
\usepackage{amsmath}
\usepackage{rotating}
\shorttitle{Statistical Study of Chromospheric Evaporation}
\shortauthors{Sadykov et al.}

\begin{document}
	
\title{Statistical Study of Chromospheric Evaporation in Impulsive Phase of Solar Flares}

\correspondingauthor{Viacheslav M Sadykov}
\email{vsadykov@njit.edu}

\author{Viacheslav M Sadykov}
\affiliation{Center for Computational Heliophysics, New Jersey Institute of Technology, Newark, NJ 07102, USA}
\affil{Department of Physics, New Jersey Institute of Technology, Newark, NJ 07102, USA}

\author{Alexander G Kosovichev}
\affiliation{Center for Computational Heliophysics, New Jersey Institute of Technology, Newark, NJ 07102, USA}
\affiliation{Department of Physics, New Jersey Institute of Technology, Newark, NJ 07102, USA}
\affiliation{NASA Ames Research Center, Moffett Field, CA 94035, USA}

\author{Ivan N Sharykin}
\affiliation{Department of Space Plasma Physics, Space Research Institute of RAS, Moscow, 117997, Russia}

\author{Graham S Kerr}
\affiliation{NASA Goddard Space Flight Center, Heliophysics Sciences Division, Code 671, Greenbelt MD 20771, USA}

\begin{abstract}
	We present a statistical study of chromospheric evaporation in solar flares using simultaneous observations by the RHESSI X-ray telescope and the IRIS UV spectrograph. The results are compared with radiation hydrodynamic flare models from the F-CHROMA RADYN database. For each event, we study spatially-resolved Doppler shifts of spectral lines formed in the transition region (C\,II\,1334.5\,{\AA}) and hot coronal plasma (Fe\,XXI\,1354.1\,{\AA}) to investigate the dynamics of the solar atmosphere during the flare impulsive phase. We estimate the energy fluxes deposited by high-energy electrons using X-ray imaging spectroscopy and assuming the standard thick-target model. Using the RADYN flare models, the RH 1.5D radiative transfer code and the Chianti atomic line database, we calculate C\,II and Fe\,XXI line profiles and compare with the observations. While the RADYN models predict a correlation between the Doppler shifts and deposited energy flux for both lines, this was only observed in the C II data. Several quantitative discrepancies are found between the observations and models: the Fe\,XXI Doppler shifts are substantially stronger in the models than in the data; the C\,II mean blueshifts are absent in the observations but predicted by the models. The transition energy between ``gentle'' and ``explosive'' evaporation regimes estimated from the observations ($2-8\times{}10^{9}$~erg~cm$^{-2}$~s$^{-1}$) and derived from the models ($2.2-10.1\times{}10^{9}$~erg~cm$^{-2}$~s$^{-1}$) are comparable with each other. The results illustrate relationships among the processes of chromospheric evaporation, response of the colder layers, and the flare energy flux deposited by high-energy electrons, although demonstrating discrepancy between analyzed observations and RADYN models.
\end{abstract}


\section{Introduction}
\label{section1}

	Solar flares are a very complicated phenomenon because of a great variety of physical mechanisms and effects simultaneously taking place in the solar atmosphere. The tremendous amount of energy released in solar flares make these events of special interest for study. The CSHKP model \citep{Carmichael64,Sturrock68,Hirayama74,Kopp76,Priest02,Shibata11} predicts that charged particles accelerated in a coronal current sheet form a particle beam that propagates into denser layers of the solar atmosphere. The beam loses its energy and momentum in the dense plasma producing a region of strong pressure and high temperature. The heated chromospheric plasma expands up to the corona along magnetic field loops. This process is called ``chromospheric evaporation''. Other mechanisms of energy deposit, such as heat conduction \citep{Antiochos78} and dissipation of high frequency Alfv\'enic waves \citep{Fletcher08,Reep16a,Kerr16}, can also initiate the chromospheric evaporation and affect its properties.
	
	Depending on the deposited energy flux, the chromospheric evaporation can be of two types: ``gentle'' (with subsonic velocities of the evaporated plasma and timescales of several minutes) driven by low-energy flux electron beams or heat conduction, and ``explosive'' (characterized by supersonic upflows on shorter timescales) driven by high-intensity electron beams \citep{Antiochos78,Zarro88}. Interestingly, the response of the chromospheric and lower transition region layers differs in these regimes. For the ``explosive'' type, the model predicts downward expansion accompanied by formation of a radiative shock and a relatively low-temperature ($\sim{}10^{4}$\,K) dense layer formed in the shock relaxation zone \citep{Kosovichev86}, which is called the ``chromospheric condensation''. For the ``gentle'' type, the downward expansion is weak, so that upward motions dominate. Numerical simulations by \citet{Fisher85a,Fisher85b,Fisher85c} confirmed the existence of such evaporation regimes, and showed that the critical energy flux for transition from the ``gentle'' to ``explosive'' is $\sim$10$^{10}$~erg~cm$^{-2}$~s$^{-1}$ for electron beams with a 20\,keV low-energy cutoff.  It is obvious that evaporation properties are closely connected to the energy release and transport mechanisms.
	
	Multiwavelength spectroscopy is of special interest for studying chromospheric evaporation because different spectral lines are formed at different heights and can represent local properties of the plasma at these heights. An overview of multiwavelength spectroscopic studies of solar flares was presented by~\citet{Milligan15}. Previous observational studies \citep[][etc]{Milligan09,Raftery09,Brosius13,Doschek13,Gomory16} have confirmed the existence of the ``gentle'' \citep{Milligan06,Sadykov15} and ``explosive'' \citep{Brosius04,Brosius15,Li15a,Brosius17} chromospheric evaporation regimes which was concluded from the observations of the Doppler shifts of the chromospheric and transition region lines of different signs. In addition, transition between these regimes had been observed during some flares \citep{Raftery09,Li15b}.
	
	Observations by the Interface Region Imaging Spectrograph \citep[IRIS,][]{DePontieu14} 
	provide a unique opportunity for detailed spectroscopic studies of dynamics of the solar atmosphere associated with chromospheric evaporation. IRIS observes several lines formed in the chromosphere and lower transition region (Mg\,II\,h\,\&\,k 2796\,{\AA} and 2803\,{\AA} lines, C\,II 1334\,{\AA} and 1335\,{\AA} lines and Si\,IV 1394\,{\AA} and 1403\,{\AA} lines), and, in addition, the Fe\,XXI\,1354.1\,{\AA} line which corresponds to a forbidden magneto-dipole transition and is formed in a very hot $\sim{}10^{7}$\,K plasma. There have been many works on analysis of the chromospheric evaporation process using the high spatial, spectroscopic and temporal resolution observations from IRIS \citep[][etc]{Tian14,Brosius15,Graham15,Kerr15,Li15b,Polito15,Sadykov15,Tian15,Young15,Polito16,Sadykov16,Zhang16,Brosius17,Lee17,Li17a,Li17b,Gupta18}. A wide range of velocities of the hot evaporated plasma and various dynamical responses of colder chromospheric layers have been reported.
	
	Radiative hydrodynamic flare simulations developed in recent years allow us to understand a complicated physics behind the observed phenomena. Many numerical simulations of the chromospheric evaporation process have been performed, both for general studies of the flare dynamics and applications to specific flare events, considering various heating mechanisms and energy release rates \citep{Kostyuk74,Livshits83,Fisher85a,Fisher85b,Fisher85c,Kosovichev86,Liu09,Brannon14,Kennedy15,Reep15,RubioDaCosta15a,RubioDaCosta15b,Kerr16,Reep16a,Reep16b,RubioDaCosta16,Johnston17a,Johnston17b}. In the simulations, it is possible to test a variety of initial conditions and heating mechanisms, understand how the atmosphere responds from the physical point of view, and derive relations between the observed characteristics of the chromospheric evaporation and the deposited energy flux and other parameters. Currently one of the most advanced code for the modeling is the RADYN radiative hydrodynamic code \citep{Carlsson97,Abbett98,Abbett99,Allred05,Allred06,Liu09,Allred15}. A grid of RADYN models is available online from the F-CHROMA project (\url{http://www.fchroma.org/}), allowing us to investigate various regimes of flare energy release in the form of non-thermal electrons, and to compare the atmospheric response to observations.
	
	The goal of this paper is to perform a statistical analysis of chromospheric evaporation in flares simultaneously observed by RHESSI and IRIS, and compare the derived relations with those obtained from the RADYN models. In particular, we focus on analysis of the C\,II\,1334.5\,{\AA} and Fe\,XXI\,1354.1\,{\AA} lines observed by IRIS in the fast scanning regimes. The synthetic line profiles are calculated using the RH\,1.5D radiative transfer program \citep{Rybicki91,Rybicki92,Uitenbroek01,Pereira15} and the Chianti atomic line database \citep{Landi13}. Section~\ref{section2} explains details of the spectroscopic data analysis and calculation of the synthetic line profiles. The analysis results are described in Section~\ref{section3}. A discussion is presented in Section~\ref{section4}, followed by a summary and conclusion in Section~\ref{section5}.

\section{Data Selection and Analysis}
\label{section2}

	\subsection{Selection of events}
	
		Using the Interactive Multi-Instrument Database of Solar Flares~\citep[IMIDSF, \url{https://helioportal.nas.nasa.gov/},][]{Sadykov17}, we select flares the impulsive phase of which was simultaneously observed by RHESSI and IRIS. Our initial selection is restricted to the flare events from the GOES and RHESSI flare lists, which have the GOES X-ray class of C5.0 or greater, located not farther than 750$^{\prime{}\prime{}}$ from the solar disk center (to avoid strong projection effects), and observed by IRIS in one of the fast scanning regimes (with $\geq$4 slit positions across the flare region covering $\geq$6$^{\prime{}\prime{}}$ in the direction perpendicular to the slit, with $\leq$90\,s cadence per cycle). The events with HXR sources not covered by the IRIS slit or with no prominent non-thermal component in HXR spectra are excluded from the analysis. Also, the events, for which the emission of Fe\,XXI\,1354.1\,{\AA} line is non-detectable by our methods, are also eliminated from the analysis. The final selection includes 7 flare events (see Table~\ref{table1_observations_viscs}) that satisfy all the criteria.

	\subsection{RHESSI data analysis}
	
		The X-ray spectroscopic data obtained by RHESSI~\citep{Lin02} allow us to estimate the deposited energy rates as well as the size and location of the hard X-ray (HXR) source and therefore derive the energy flux, one of the key parameters for the flare hydrodynamic simulations.
		
		To estimate the deposited energy rate we fit the thermal plus non-thermal thick-target model~\citep{Brown71,Kostiuk75} to the X-ray spectra calculated for five 12\,sec intervals (or 18-24\,s for the flares with low count rates) covering the emission peaks in the 25-50\,keV energy range. The pileup correction and isotropic albedo component are applied to the spectra. The total time intervals, fitting energy intervals, and the RHESSI detectors used for our spectral analysis are listed in Table~\ref{table1_observations_viscs}. In the case of several HXR peaks we select the one preceding the enhancement of UV line characteristics observed by the IRIS satellite.
		
		The fitting is performed using a least-squares procedure available from the OSPEX Solar SoftWare \citep[SSW,][]{Freeland00} package. The fit functions include three components: ``vth'', isotropic ``albedo'', and ``thick2''. The deposited energy rate for each time interval is calculated using the formula $P_{\textrm{nonth}} = \dfrac{\delta{}-1}{\delta{}-2}F(E>E_{\textrm{c}})E_{\textrm{c}}$, where $E_{\textrm{c}}$ is the low-energy cutoff derived from the spectra, $F(E>E_{\textrm{c}})$ is the number of deposited electrons per second, $\delta{}$ is the spectral index. From the five time intervals for each flare we select two intervals with prominent non-thermal components, the highest deposited energy rates, and the smallest relative errors of the fitting parameters. In addition to the deposited energy rates, for each event we derive the averaged parameters of the energy spectra, $\delta{}$ and $E_{\textrm{c}}$. We note here that $E_{\textrm{c}}$ is notoriously hard to measure, and consider a maximum value that is consistent with the data, meaning that the derived power carried by non-thermal electrons is a lower limit.
		
		To determine the energy flux, we reconstruct the RHESSI images using the recently developed ``vis\_cs'' algorithm~\citep{Felix17} applied with the standard $\lambda{}=0.5$ sparseness parameter to the entire time intervals of the RHESSI analysis. The data from detectors 2-8 are used for image reconstruction. The source size, S, is determined as the area within 50\% contours of the reconstructed images of the HXR flux in the energy range of 25-50\,keV. The reconstructed HXR sources areas are likely overestimated, meaning that the stated energy fluxes are lower limits. Finally, we derive the flux rate of non-thermal electrons for each flare event as $F_{\textrm{nonth}} = P_{\textrm{nonth}}/S$ (projection effects are taken into account in the source size calculation) and use this value as the deposited energy flux. Along with the low-energy cutoff, $E_{\textrm{c}}$, and the power-law index, $\delta{}$, it is used for identifying the closest RADYN model for the analyzed flares. In addition, we calculate the mean 25-50\,keV photon fluxes emitted from the HXR sources by dividing the integrated flux by the source area.

	\subsection{IRIS data analysis}
	
		The spatially-resolved measurements of ultraviolet (UV) spectra obtained by IRIS \citep{DePontieu14} allow us to understand the properties of the evaporated plasma, as well as the dynamical response of the lower layers of the solar atmosphere. In this study, we focus on analysis of the C\,II\,1334.5\,{\AA} ($1-2\cdot{}10^{4}$\,K) and Fe\,XXI\,1354.1\,{\AA} ($\approx$10$^{7}$\,K) lines which reflect the dynamics of the relatively cold chromospheric and lower transition region layers (including the chromospheric condensation) and the hot evaporating plasma respectively. We choose the C\,II line because it has a less complex profile than the Mg\,II lines, and it is rarely overexposed during the flares. However, the C\,II line is still optically-thick \citep{Rathore15} which can make any interpretation of C\,II Doppler shifts ambiguous~\citep{Kuridze15}.
		
\begin{figure}
	\centering
	\includegraphics[width=1.0\linewidth]{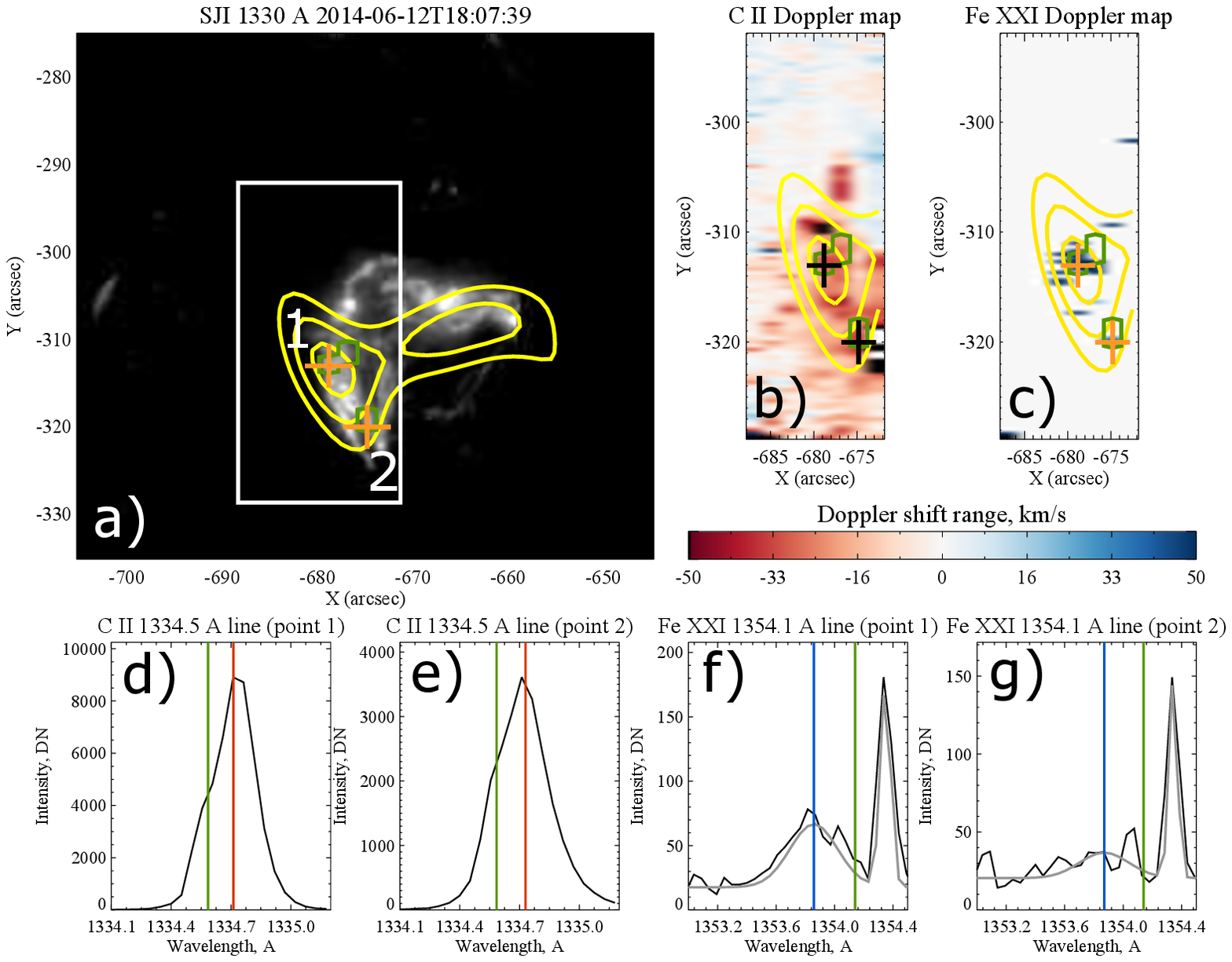}
	\caption{An example of joint IRIS and RHESSI observations of the SOL2014-06-12T18:03:00 flare event. a) IRIS 1330\,{\AA} slit-jaw image. White rectangle corresponds to the region scanned by the spectrograph slit. b) Map of C\,II\,1334.5\,{\AA} line Doppler shifts. c) Map of Fe\,XXI\,{1354.1}\,{\AA} line Doppler shifts. Yellow contours in panels a)-c) correspond to the 50\%, 70\%, and 90\% of the maximum of the HXR 25-50\,keV source reconstructed from RHESSI data for 18:07:30-18:08:30 time period. Green contours correspond to the mask of points for which the Doppler estimates (mean and strongest) are obtained. d,e) C\,II line profiles obtained for points 1 and 2 (orange and black crosses in panels a-c). Red vertical line corresponds to the centroid of the line, green vertical line~--- to its reference wavelength. f,g) Fe\,XXI and C\,II line profiles obtained for points 1 and 2 (black and grey crosses in panels a-c). Grey profiles show the corresponding double-Gaussian fitting. Blue vertical lines indicate the center of the Gaussian corresponding to the Fe\,XXI line, green vertical lines~--- its reference wavelength.}
	\label{figure1_maps}
\end{figure}

		For analysis of the C\,II line, we calculate the center-of-gravity parameters: (1)~the line peak intensity, and (2)~the Doppler shift defined as the difference between the center of gravity of the line and the reference wavelength for this line $\left<{\lambda}\right> - \lambda{}_{ref}= {\int{{\lambda}Id{\lambda}}}/{\int{Id{\lambda}}} - \lambda{}_{ref}$. This approach was tested in our previous works~\citep{Sadykov15,Sadykov16}. We calculate these characteristics for every spatial point and every available time moment, and reconstruct time-dependent maps of the line intensity and Doppler shift. Examples of such maps and the line profiles are presented in Figure~\ref{figure1_maps} for SOL2014-06-12T18:03:00 event. The reference wavelength for the C\,II line, $\lambda{}_{\textrm{ref}}$, is estimated for each flare separately from the spectra in the areas not affected by the flares.
		
		The Center-of-Gravity estimates cannot be directly applied to the Fe\,XXI line, because it is blended with other lines. The strongest blend of the Fe\,XXI line is the C\,I\,1354.334\,{\AA} line. These two lines are dominant in the corresponding IRIS spectral window during the flares. Thus, we fit the spectra using the double-Gaussian fitting, the applicability of which was demonstrated by~\citet{Battaglia15}. From the fitting parameters, we estimate the peak intensity of the Fe\,XXI line (as the amplitude of the corresponding Gaussian) and its Doppler shift (as a mean shift of the corresponding Gaussian from the reference wavelength). The spectra where the fitting results are unreliable (the intensity in the channel does not exceed preflare activity level, the standard deviation of the Gaussian corresponding to C\,I line exceeds 0.15\,{\AA}, and the standard deviation of the Gaussian corresponding to the Fe\,XXI line is outside the 0.15\,{\AA}-0.30\,{\AA} range) are not considered in the analysis. The reference wavelength of the Fe\,XXI line is kept equal to 1354.14\,{\AA} as in our previous work~\citep{Sadykov15}. This estimate deviates from the value of 1354.106\,{\AA} derived by~\citet{Young15} by $\sim{}$0.03\,{\AA} (or $\sim$6\,km/s). However, the difference does not affect analysis of chromospheric evaporation flows with velocities of $\sim$100\,km/s or greater. An example of Fe\,XXI Doppler shift map together with examples of Fe\,XXI and C\,I line profiles and double-Gaussian fitting are also presented in Figure~\ref{figure1_maps}.
		
		After calculating the line intensities and Doppler shifts for the entire flare region, we determine a time-dependent mask of points involved in the flare heating. First, we find the maximum of the averaged C\,II line intensities (integrated across the scanned region) for each scan during the flare and define it as a threshold, and then for each scan, construct the mask of points where the C\,II intensity is greater than the threshold. The derived masks capture the flare dynamics and are independent from the RHESSI data. A comparison of the mask derived for SOL2014-06-12T18:03:00 with the HXR sources is presented in Figure~\ref{figure1_maps}a-c. Basically, for each considered flare event the derived IRIS masks intersect the previously obtained regions of the 25-50\,keV HXR sources.
		
\begin{figure}
	\centering
	\includegraphics[width=1.0\linewidth]{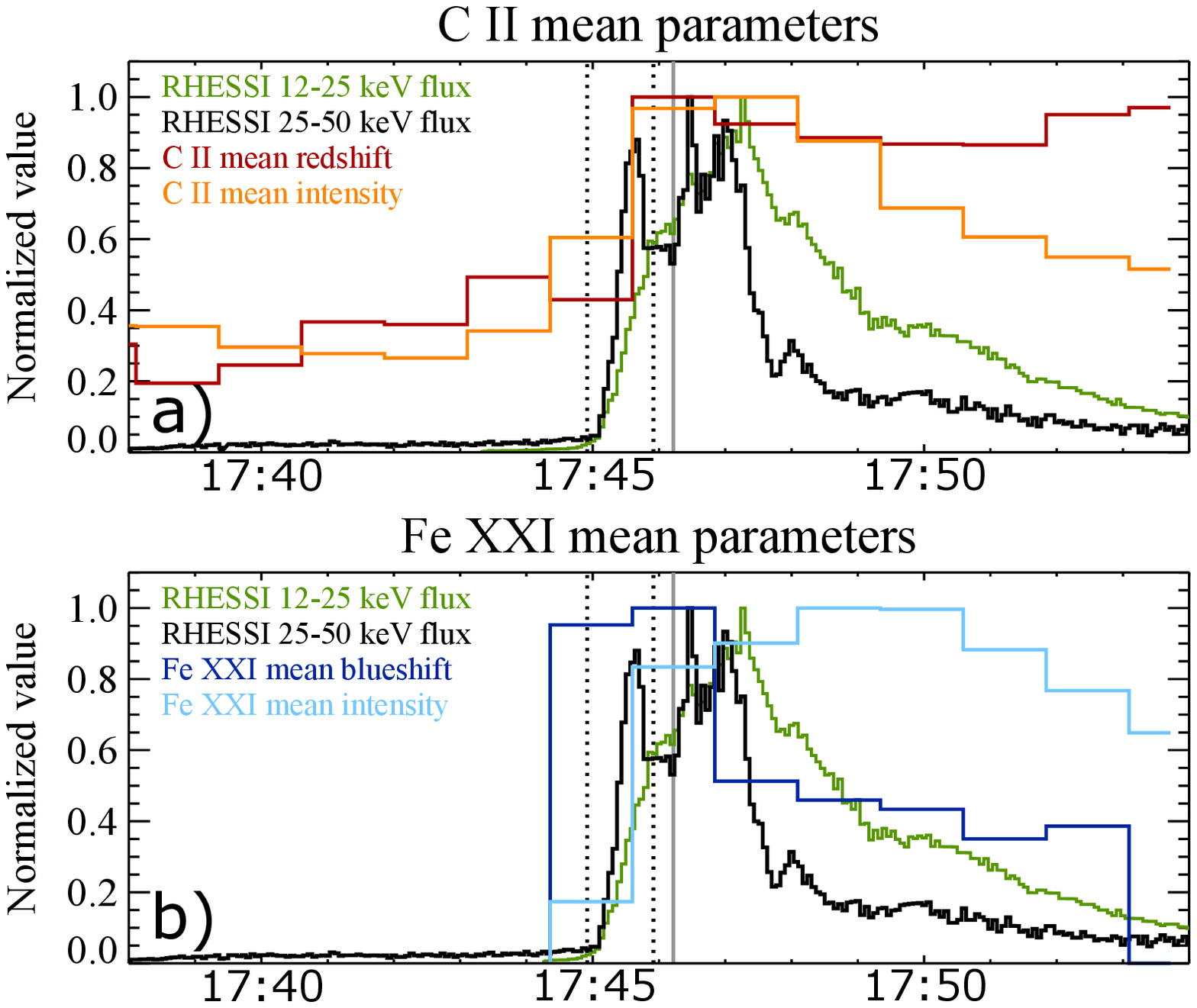}
	\caption{An example of the normalized integrated light curves and variations of the spectral line parameters for the SOL2014-03-29T17:35:00 flare event. a) normalized HXR 12-25\,keV (green), 25-50\,keV (black) light curves and normalized C\,II\,1334.5\,{\AA} intensity (orange) and Doppler shift (red). b) normalized HXR 12-25\,keV (green), 25-50\,keV (black) light curves and normalized Fe\,XXI\,1354.1\,{\AA} intensity (light blue) and Doppler shift (dark blue). Black dotted vertical lines mark the time interval where the deposited energy flux was estimated from RHESSI data. Grey vertical line represents the middle of the time interval where the peaks of the Doppler shifts means occur.}
	\label{figure2_gencurve}
\end{figure}

		To quantify the response of the solar atmosphere, we calculate the mean values of the C\,II and Fe\,XXI Doppler shifts within the derived masks, and record the time and value of their maxima. We calculate the standard deviation for the mean Doppler shift values to estimate their uncertainties. An example of the evolution of C\,II and Fe\,XXI mean Doppler shifts during the SOL2014-03-29T17:35:00 X1.0 class flare is presented in Figure~\ref{figure2_gencurve}. The mean Doppler shift peaks are very prominent, and correspond to the first pulse of HXR 25-50\,keV flux. In addition, we quantify the Doppler shifts in the areas that responded most strongly to the heating. The following procedure is applied: 1) for each IRIS slit scan, derive the Doppler shift level above which 95\% of detected C\,II Doppler shifts are located (i.e. separating the 5\% strongest C\,II redshifts across the flare region); 2) find the minimum among the derived Doppler shift levels. We repeat the same procedure for the Fe\,XXI blueshifts (hereafter referred as ``Fe\,XXI strongest blueshifts''). Their values and times are summarized in Table~\ref{table1_observations_viscs} for each flare. To estimate errors in the determination of the C\,II strongest redshifts and Fe\,XXI strongest blueshifts, we decreased the IRIS raster field-of-view by 25\% and repeated the procedure. The standard deviations for the mean Doppler shifts are comparable with the Doppler shifts themselves, but the uncertainties for the strongest Doppler shift values are usually lower than the IRIS spectral resolution value, $\approx$3\,km/s, for the C\,II line, and the uncertainty in the reference wavelength of the Fe\,XXI line.

	\subsection{Calculation of synthetic line profiles}
	
		The F-CHROMA database provides the 1D radiative hydrodynamic (RADYN) models of solar flares for a variety of the electron beam parameters (averaged energy fluxes from 1.5$\times$10$^{9}$ to 5.0$\times$10$^{10}$\,erg\,cm$^{-2}$s$^{-1}$, low-energy cutoff values of 15\,keV, 20\,keV, or 25\,keV, and spectral indexes ranging from 3 to 8). The RADYN code solves the coupled, non-linear, equations of hydrodynamics, radiation transport, and non-equilibrium atomic level populations, on an adaptive 1D vertical grid. The elements that are important for the chromospheric energy balance are treated in the non-LTE formulation, and the other species are included in the radiative loss function in the LTE approximation. The atomic level population and radiation transport equations are solved for a 6-level-with-continuum hydrogen atom, a 9-level-with-continuum helium atom, and a 6-level-with-continuum Ca\,II atom. For a detailed description see \citet{Allred15} and references therein. In the F-CHROMA database, the 1D flare models are calculated with 300 height grid points and 201 frequency points. To avoid overestimating radiative losses from the Ly-$\alpha{}$ line partial redistribution (PRD) effects were mimicked by modeling these lines as Doppler profiles. The initial atmosphere is described using the VAL3C model \citep{Vernazza81}. The temporal profile of the deposited energy flux rate is a triangle; the electron beam heating lasts for 20\,s with the peak at 10\,s. In this work, we analyze 20 RADYN models. Although there are models which are close to the analyzed flares in terms of the averaged energy flux, averaged low-energy cutoff and spectral index, we do not explicitly compare them to each other because of ambiguities (and possible overestimations) in HXR source areas derived from RHESSI data. The selection results are summarized in Table~\ref{table2_modeling}.
		
		To calculate the C\,II\,1334.5\,{\AA} line we use the height scale, density, temperature, electron density, plasma vertical velocity, and hydrogen populations from the RADYN snapshots as an input for the RH radiative transfer code~\citep{Rybicki91,Rybicki92,Uitenbroek01,Pereira15}. The latest version of the RH 1.5D massively-parallel code (\url{https://github.com/ITA-Solar/rh}) has been adopted for the calculations. Notice that the code assumes that the populations of atomic levels are in the statistical equilibrium, but the non-equilibrium electron density distribution is taken from the RADYN models. The inclusion of the non-LTE hydrogen, carbon, and silicon \citep[an important source of background opacity for C\,II lines,][]{Rathore15,Kerr17a} populations provides results accurate enough for the C\,II line profile calculations, other species are assumed to be in LTE. Also, the calculations are done under the complete frequency redistribution assumption (CRD) that has been proven to be adequate for the C\,II line \citep{Rathore15}. The C\,II line profiles are calculated with 1\,s time cadence for the selected RADYN models.
		
		To calculate the Fe\,XXI\,1354.1\,{\AA} line, we use the Chianti atomic line database~\citep{Landi13}. It allows us to simulate the optically-thin emission of the Fe\,XXI line under the ionization equilibrium assumption for a single temperature plasma. Using the Chianti software we simulate the Fe\,XXI emission at each grid point of the RADYN model (assuming thermal line broadening only), calculate the emission Doppler shift according to the local plasma velocities, and sum up the results for each snapshot.
		
\begin{figure}
	\centering
	\includegraphics[width=1.0\linewidth]{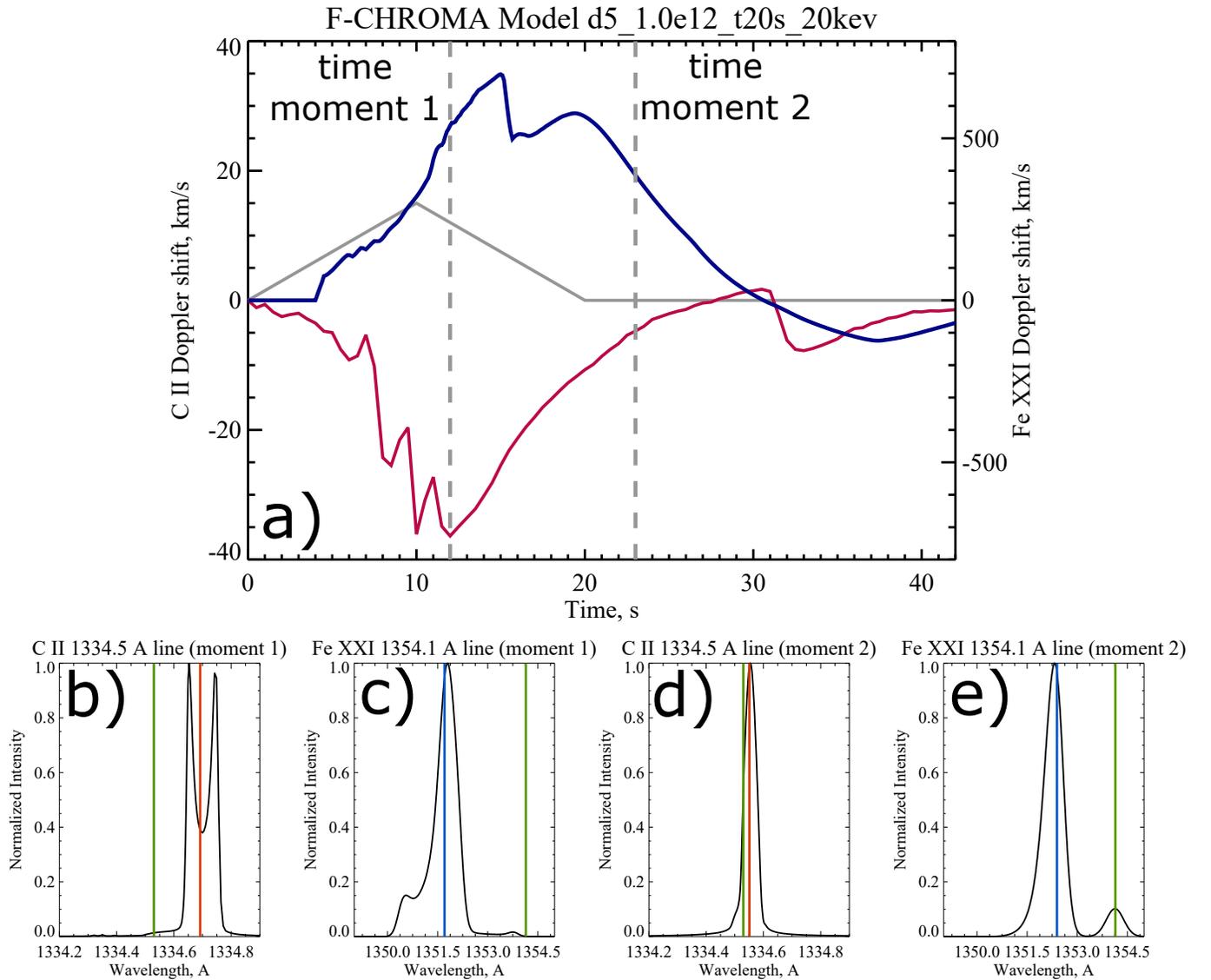}
	\caption{a) An example of the C\,II\,1334.5\,{\AA} and Fe\,XXI\,1354.1\,{\AA} line Doppler shift calculations for the RADYN run d5\_1.0e12\_t20s\_20keV. In this simulation, a non-thermal electron distribution with a slope of $\delta = 5$, low energy cutoff $E_{\textrm{c}} = 20$~keV was injected for 20s, according to the triangle-shaped profile (gray), delivering a total time-integrated flux of 1.0 $\times{} 10^{12}$~erg~cm$^{-2}$. The corresponding C\,II (red, left scale) and Fe\,XXI (blue, right scale) Doppler shift evolutions are presented. Gray dashed lines mark the time moments for which the simulated C\,II and Fe\,XXI line profiles in panels b)-e) are presented. Red and blue vertical lines in panels b)-e) correspond to the centroids of the lines, green vertical lines~--- to their reference wavelengths.}
	\label{figure3_radynexample}
\end{figure}

		Finally, the Doppler shift is calculated using the Center-of-Gravity method for both C\,II and Fe\,XXI lines. An example of the Doppler shift behavior and the simulated C\,II and Fe\,XXI line profiles is presented in Figure~\ref{figure3_radynexample}, which illustrates the ``explosive'' chromospheric evaporation (with strong redshifts of the C\,II line). For each run, we record the peak C\,II Doppler shifts during or within five seconds after the heating phase, and the peak Fe\,XXI Doppler shifts during the entire run, and use them as a measure of the atmospheric response to the heating. The peak values of the Doppler shifts for each model are shown in Table~\ref{table2_modeling}. The missing Fe\,XXI values correspond to the RADYN models for which the plasma temperature does not exceed $10^{6}$\,K or the Fe\,XXI Doppler shift was still increasing in the end of the model. We do not apply any instrumental effects to our synthetic spectra (resolution, point-spread-function). Also, we assume that the emergent synthetic spectra are at disk center, originating from a vertical flux tube. While this is an adequate assumption for C\,II line which forms over in a narrow region deep in the flare loop, Fe\,XXI line likely forms over an extended region. Therefore when we sum up the emission from Fe XXI we are not separating footpoint emission from looptop emission. Our observations originate from various locations on the disk, and so geometric effects in some events may account for some differences between our model-data comparison.

\section{Results}
\label{section3}

	In this section, we analyze correlations among the observed flare parameters and compare them with the correlations found for the RADYN RHD flare models. Relationships between the energy fluxes and Doppler shifts are of particular interest. Such relations can provide a possibility to diagnose the properties of the energy release from the observed UV spectroscopic data. To analyze presence of correlations for each considered pair of parameters we calculate non-parametric Kendall's $\tau$ coefficient (Kendall's rank correlation coefficient) defined as:
	
	\begin{gather}
	\label{eq:Kendallstau}
	\tau{} = \dfrac{2}{n(n-1)}\sum_{i<j}sgn(x_{i}-x_{j})sgn(y_{i}-y_{j})
	\end{gather}
	
	Here $\{x_{i}\}$ and $\{y_{i}\}$ are the values of the considered pair of parameter; $sgn$ is a sign operator; $n$ is a number of elements in each data set. Kendall's $\tau$ ranges between -1 and 1, and its value is expected to be 0 for independent data sets. We calculate a p-value for a hypothesis test whose null hypothesis is an absence of association ($\tau{}=0$). Low p-value ($<$ 0.05) indicates that the difference of the presented $\tau$ values from zero is statistically significant. We also calculate linear regression correlation coefficients (CC) and the corresponding p-values (for a hypothesis test whose null hypothesis is that the slope is zero), with a weighted least squares procedure applied for the regression. Results of the statistical analysis, together with the empirical dependences for parameters, are summarized in Table~\ref{table3_regression}.
	
	Figure~\ref{figure4_photonflux_electronflux} presents the relationship between the 25-50\,keV averaged photon flux and the estimated electron energy flux for the analyzed flare events. The correlation between these parameters in the log-log scale is significant ($\tau$=0.90, p-value 0.0043), with the linear correlation coefficient (CC) of 0.88. The observed correlation confirms the applicability of the thick-target flare model which assumes that the observed HXR emission is a bremsstrahlung radiation of high-energy electrons.
	
	Figure~\ref{figure5_eflux_ciired} shows the observed relationships between (a) the C\,II\,1334.5\,{\AA} line mean Doppler shift and the deposited energy flux, and (b) the C\,II line strongest redshifts and the deposited energy flux. Panel (c) shows the C\,II Doppler shift vs the energy flux for the RADYN models. Although all panels demonstrate correlations between the deposited energy flux and the C\,II Doppler shift parameters (with $\tau$ of -0.52, -0.52, and -0.58, and with linear CCs of -0.42, -0.67, and -0.73 for panels a-c), the correlations are statistically significant only for models (p-values for observational relationships are high). However, the correlation between the C\,II strongest redshifts and the deposited energy flux also has a trend toward significance: the corresponding p-value for Kendall's $\tau$ is $\sim{}$0.1. The strongest difference between the models and observations (panels a and b) is that no positive C\,II mean Doppler shifts (blueshifts) are found in the observations, although according to the RADYN models these are expected for some of the observed flare events.
	
	Figure~\ref{figure6_eflux_fexxiblue} shows the observed relationship between (a) the Fe\,XXI,1354.1\,{\AA} line mean Doppler shift and the deposited energy flux, and (b) the Fe\,XXI line strongest blueshifts and the deposited energy flux. Panel (c) shows the Fe\,XXI Doppler shift and energy flux relationships for the RADYN models. Although the models demonstrate a very strong correlation (panel c, $\tau{}=0.77$, CC $=$ 0.84), the observed Fe\,XXI Doppler shift mean values do not show any significant dependence on the deposited energy flux. The Fe\,XXI strongest blueshifts show weak correlation ($\tau$=0.33, CC=0.39) with the energy flux which, moreover, cannot be called statistically-significant. We still show the corresponding linear fit in Fig.~\ref{figure6_eflux_fexxiblue}b, however, point out that low number of events does not allow us to confirm or decline presented trend. Also, one can see that many models demonstrate very strong Fe\,XXI Doppler shifts, in the range of 400-700\,km/s, while there is only one observational result exceeding 200\,km/s.
	
	Figure~\ref{figure7_fexxiblue_ciired} shows the observed relationship between (a) the C\,II\,1334.5\,{\AA} line mean Doppler shift and the Fe\,XXI,1354.1\,{\AA} line mean Doppler shift, and (b) the C\,II line strongest redshifts and Fe\,XXI strongest blueshift. Panel (c) shows the C\,II Doppler shift and Fe\,XXI Doppler shift relationships for the corresponding RADYN models. Although the correlations for the models are strong demonstrating that higher velocities of chromospheric evaporation correspond to faster downflows in the colder atmospheric layers, we do not find such behavior for observations.

\section{Discussion}
\label{section4}

	In this work, we have performed a statistical analysis of the Doppler shifts of two UV lines, C\,II\,1334.5\,{\AA} and Fe\,XXI\,1354.1\,{\AA}, which characterize the lower transition region and coronal dynamics during solar flares, and compared the results with the corresponding radiative hydrodynamic RADYN models. To estimate the deposited energy flux, we assumed the thick-target flare model of bremsstrahlung radiation. Figure~\ref{figure4_photonflux_electronflux} illustrates the general correctness of this assumption. Such kind of relationships potentially allows us to filter out the flare events for which the thick-target model is inappropriate or the fitting procedure is performed incorrectly.
	
	It is found from the RADYN flare hydrodynamic models of chromospheric evaporation that the relationships between the Doppler shifts of the UV lines and the energy flux (Figures~\ref{figure5_eflux_ciired}c-\ref{figure7_fexxiblue_ciired}c) can be approximated by a linear-log regression (see the correlation coefficients and fitting parameters in Table~\ref{table3_regression}), although the considered data set has a multi-parametric nature (low-energy cutoffs and slopes of the energetic electron spectra are not taken into account in the correlation analysis). Thus, one should expect to find similar trends for the observations to reasonable degree, given the assumptions of the model, the lack of geometric considerations, and the impact of spatio-temporal resolution on the observed profiles. However, such trends are not so clear for the considered observational data set. Figures~\ref{figure5_eflux_ciired}a,b-\ref{figure7_fexxiblue_ciired}a,b show significantly weaker correlations. The only correlation with a tendency to be statistically-significant is found between the C\,II strongest redshift and the deposited energy flux, and is described by the empirical Eq.~3 (Table~\ref{table3_regression}). In principle, this relation could be used as an indirect diagnostic tool of the deposited energy flux, allowing the estimation of the energy flux at least by an order of magnitude. We cannot make a comparison with the modeling Eq.~5 due to absence of the mean blueshifts in observational results.
	
	In Section~\ref{section1} we have discussed two regimes of chromospheric evaporation: the ``explosive'' regime characterized by supersonic velocities of evaporated plasma and chromospheric condensation downflow, and ``gentle'' regime with subsonic evaporation and upflows of colder layers. Using the empirical relations we can estimate the energy flux corresponding to transition between these regimes. For RADYN models, we use Eq.~5~and~6 in Table~\ref{table3_regression} assuming $v^{\textrm{C\,II}}[\textrm{km/s}] = 0$ and the coronal sound speed $v^{\textrm{Fe\,XXI}} = $100-200\,km/s. Considering the uncertainties of fitting coefficients, we obtain that the transition energy flux is (2.2-10.1)$\cdot{}$10$^{9}$~erg~cm$^{-2}$~s$^{-1}$ for the models. We note that this flux is lower than the $\sim$10$^{10}$~erg~cm$^{-2}$~s$^{-1}$ transition flux obtained by \citet[][20\,keV low-energy cutoff is assumed for the beam]{Fisher85a}. Unfortunately, it is impossible to derive reliable transition energy flux from the observations. The only way to get an estimate is to use the Eq.~5 (Table~\ref{table3_regression}) for Fe\,XXI strongest blueshifts, because no positive mean Doppler shifts were observed for C\,II, and no correlation for the Fe\,XXI mean Doppler shift was found. Assuming that there are no supersonic flows (greater than 100-200\,km/s) for the flares with gentle evaporation, we estimate the transition energy flux to be (2-8)$\cdot{}$10$^{9}$~erg~cm$^{-2}$~s$^{-1}$. Notice again that high ambiguities of the observational relation and high p-value do not allow us to claim that the estimated transition energy flux is statistically-reliable. We also need to mention that the transition energy depends in general not only on the deposited energy fluxes, but also on the low-energy cutoff values, the power law of the non-thermal spectra, and the duration of heating \citep{Fisher89,Reep15}. It seems that the low-energy cutoff dependence dominates over the spectral slope variations \citep{Sharykin16}. We see signatures of this effect in the models.
	
	Our analysis revealed the difference between the simulated and observed Fe\,XXI Doppler velocities: the observed Doppler velocities rarely exceed 150\,km/s, while the Doppler shifts calculated for the RADYN models indicate significantly higher upflow velocities, ranging from 200\,km/s to 700\,km/s. The Doppler shifts of the Fe\,XXI line detected in other works also rarely exceed 200\,km/s. For example, \citet{Young15} reported the evaporation velocities of about 100-200 km/s for the SOL2014-03-29T17:35:00 flare. \citet{Brosius15} reported Doppler shifts of 150\,km/s for the Fe\,XXI line during the M7 flare of 2014 April 18. \citet{Polito15} detected 82\,km/s blueshift for the C6.5 class flare of 2014 February 3. For two X1.6 class solar flares of 2014 September 10 and October 22, \citet{Li15a} found the Fe\,XXI velocities up to 200\,km/s. \citet{Zhang16} found the upflows of 35-120\,km/s during the C4.2 circular-ribbon flare on 2015 October 16. \citet{Polito16} detected the evaporation flow velocities of 200\,km/s for fully-resolved (single-gaussian) Fe\,XXI line profile. Why does this discrepancy happen? \citet{Graham15} pointed out that the Fe\,XXI line is very strongly blueshifted in the beginning of the flare, but its emission is weak. As evaporation develops, the line becomes stronger but less blueshifted. This effect is also seen in Figure~\ref{figure2_gencurve}b. The weak forbidden Fe\,XXI line, with several blends on it, might simply be non-detectable during the most blueshifted phase, when the maximum of the evaporation velocities takes place. The RADYN models confirm such line behavior in general: the intensity of the synthesized Fe\,XXI line continues to grow at the time of the Doppler shift peak for most of the models. We also note here that it is not possible to detect Fe\,XXI line blueshifts of $\gtrapprox{}$300\,km/s because of the limited wavelength range of IRIS O\,I spectral channel. However, the Doppler shifts found in this work rarely exceed 200\,km/s, and there is no strong influence of this effect on our results.
	
	The RADYN models available from the F-CHROMA database (\url{http://www.fchroma.org/}) have a standard time dependence of the energy input in a form of triangle 20\,s duration with the peak at 10\,s. To estimate the influence of the deposited energy flux profile shape on the Doppler shifts of the C II and Fe XXI lines without changing the heating phase duration and energetics (average and peak deposited energy fluxes), we performed four additional RADYN runs. Two of them were similar to the run ``d5\_1.0e12\_t20s\_25keV'', with the only difference that the location of the peak of the heating function was at 5\,s (1$^{st}$ run) and at 15\,s (2$^{nd}$ run) from the start of the run. The 2$^{nd}$ run was computationally demanding, and so we show here the first 18\,s that we have computed (long enough to demonstrate any large scale differences). For two other runs, the total energy of $10^{10}$~erg~cm$^{-2}$ (lower than in any F-CHROMA run) was deposited by a population of accelerated electrons with 15\,keV low-energy cutoff and spectral slope of 7. The heating was, again, triangular shape, lasted 20\,s, and peaked at 5\,s (3$^{rd}$ run) and at 15\,s (4$^{th}$ run). We applied the same analysis for these runs as for the F-CHROMA runs, and calculated the peak Doppler shifts for C\,II and Fe\,XXI. For the 1$^{st}$ and 2$^{nd}$ runs, the C\,II peak Doppler shift was -21.6\,km/s and -13.4\,km/s, which is relatively close to the Doppler shift derived for ``d5\_1.0e12\_t20s\_25keV'' run (-15.7\,km/s). The corresponding Fe XXI peak Doppler shift for the 1$^{st}$ run was 550\,km/s (compare with 706\,km/s for ``d5\_1.0e12\_t20s\_25keV'' run). Unfortunately, the Fe\,XXI peak Doppler shift was not reached during the first 18\,s of the 2$^{nd}$ run, but the calculated Doppler velocities of the Fe\,XXI line were $>$330\,km/s at 18\,s and continued to grow. The 3$^{rd}$ and 4$^{th}$ runs showed consistent behavior. The peak Doppler shifts for these runs were 20.5\,km/s and 21.2\,km/s (for C\,II line) and 34\,km/s and 35\,km/s (for Fe\,XXI line) correspondingly. Thus, we can expect that the heating profile variations (with the total and peak deposited energy fluxes fixed) do not significantly affect the overall response and large scale trends of the atmospheric response, but do introduce differences.
	
	Among the studied flares, we did not find any with positive mean Doppler shift of the C\,II line. The C\,II line is always mainly redshifted in observations. However, the RADYN models suggest that for the low energy fluxes found in several studied events, we should detect ``gentle'' chromospheric evaporation with blueshifts of the C\,II line. One of the possibilities to explain this discrepancy is that we significantly overestimate the area into which energy is deposited, thus underestimating the energy fluxes of the observed events. As an example, the deposited energy flux for the SOL2014-03-29T17:35:00 flare estimated by~\citet{Kleint16} is more than 10 times higher than one obtained in this work. On the other hand, \citet{Sadykov15} previously found gentle chromospheric evaporation during the SOL2014-06-12T21:01:00 M1.0 where the C\,II line was also mainly redshifted. Despite negative values of the C\,II mean Doppler shifts, we found that for all observed events there are areas with blueshifted C\,II line during the impulsive phase. This finding is in agreement with the multi-thread nature of solar flares \citep{Graham15,Warren16,Reep16b,RubioDaCosta16} and requires further detailed investigation.

\section{Summary and Conclusion}
\label{section5}

\begin{figure}
	\centering
	\includegraphics[width=0.8\linewidth]{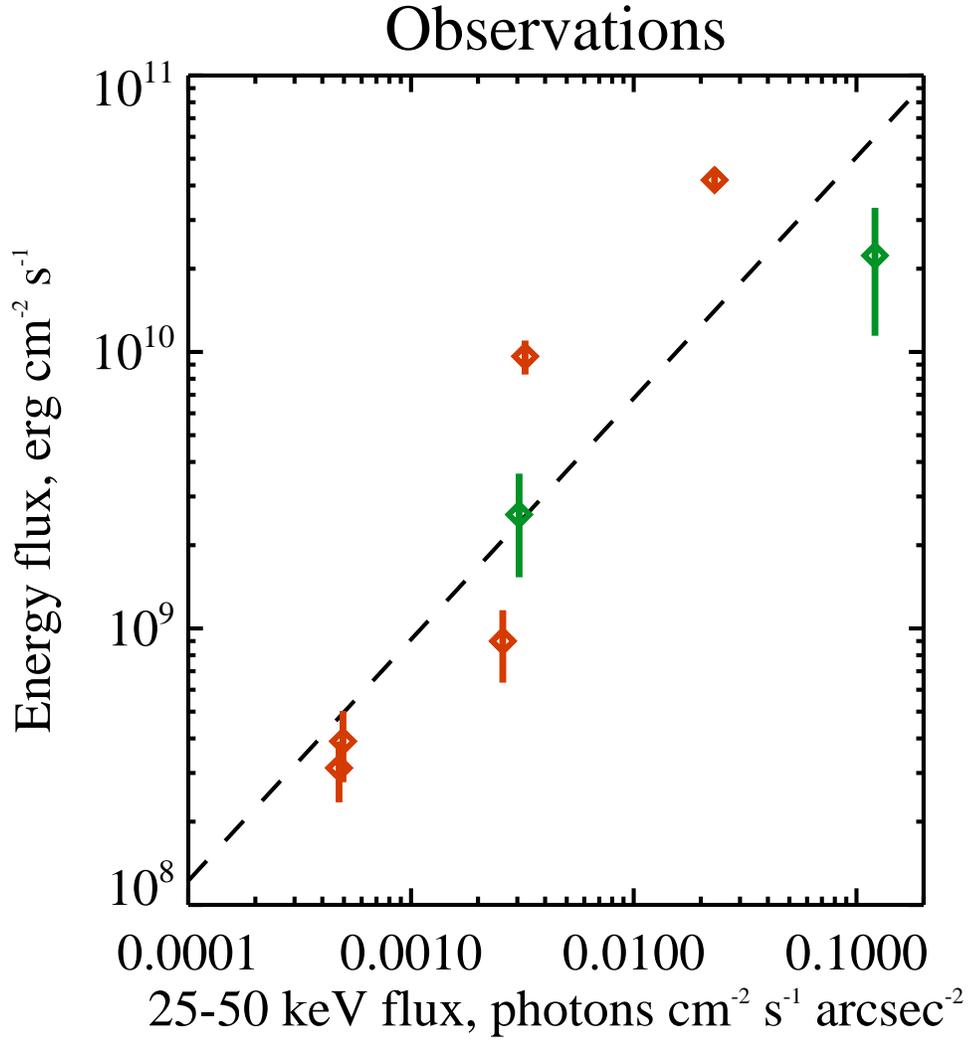}
	\caption{Dependence of the deposited electron energy flux estimated using the thick-target model fits to the X-ray spectra on the mean flux of 25-50\,keV photons for the analyzed flare events. Different colors correspond to the closest low-energy cutoff values among 15\,keV (green), 20\,keV (red) and 25\,keV (black). The inclined dashed line indicates the best linear log-log fit. The linear fit coefficients and corresponding correlation coefficient are presented in Table~\ref{table3_regression}.}
	\label{figure4_photonflux_electronflux}
\end{figure}

\begin{sidewaysfigure}
	\centering
	\includegraphics[width=1.00\linewidth]{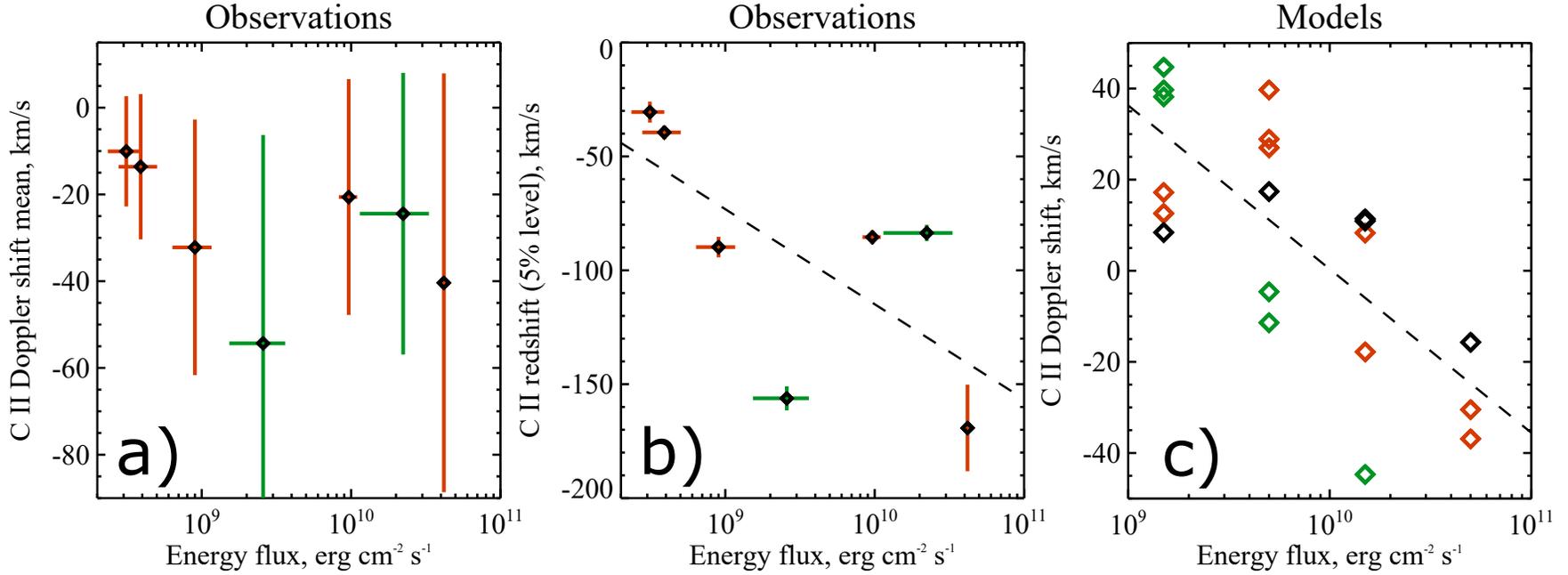}
	\caption{Dependence of the C\,II\,1334.5\,{\AA} line Doppler shift estimates on the deposited energy flux from the observations (panels a and b, mean values and the strongest redshifts) and the RADYN models (panel c, strongest Doppler shifts over the run). The size of crosses in panels b) and c) corresponds to the errors estimated for each flare event. Different colors correspond to the closest low-energy cutoff values among 15\,keV (green), 20\,keV (red) and 25\,keV (black). The inclined dashed lines show the best linear fits. The fitting and correlation coefficients are presented in Table~\ref{table3_regression}.}
	\label{figure5_eflux_ciired}
\end{sidewaysfigure}

\begin{sidewaysfigure}
	\centering
	\includegraphics[width=1.00\linewidth]{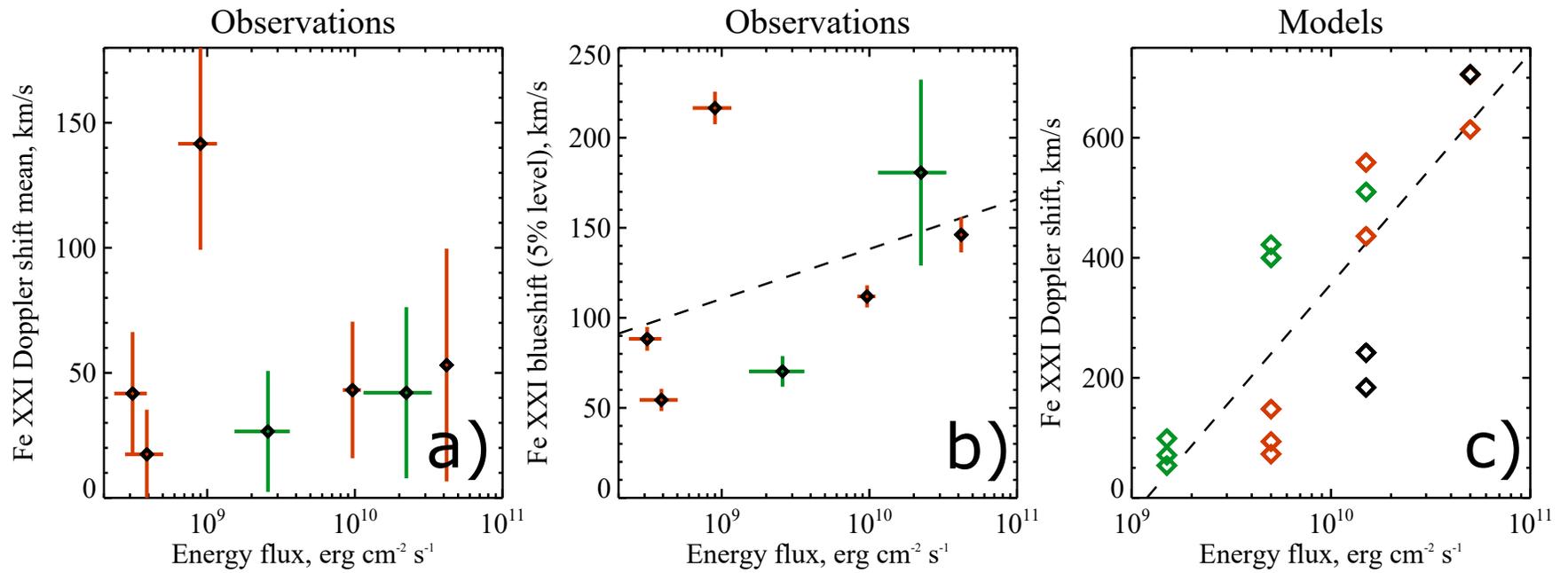}
	\caption{The same as in Fig.~\ref{figure5_eflux_ciired} for the relationship between the energy flux and the Fe\,XXI 1354.1\,{\AA} line Doppler shift.}
	\label{figure6_eflux_fexxiblue}
\end{sidewaysfigure}

\begin{sidewaysfigure}
	\centering
	\includegraphics[width=1.00\linewidth]{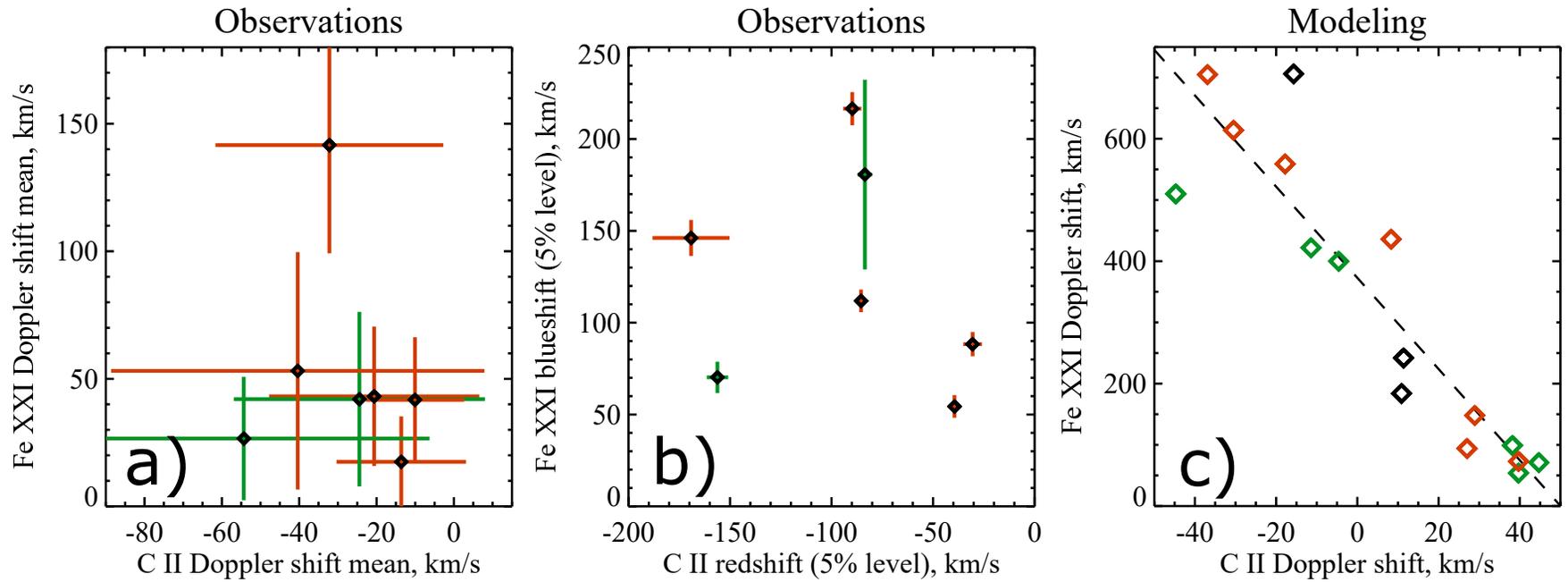}
	\caption{The same as in Fig.~\ref{figure5_eflux_ciired} for the relationship between the C\,II 1334.5\,{\AA} and Fe\,XXI 1354.1\,{\AA} line Doppler shift estimates.}
	\label{figure7_fexxiblue_ciired}
\end{sidewaysfigure}

	In summary,
	\begin{enumerate}
		\item We analyzed seven flares jointly observed by the RHESSI and IRIS satellites that allowed us to perform a statistical study of the chromospheric evaporation, and investigate relationships between the energy release properties and the atmospheric response. To compare observational findings with the results of the chromospheric evaporation process simulations, we calculated synthetic C\,II and Fe\,XXI line profiles for 20 different radiative hydrodynamic (RADYN) models and derived the corresponding Doppler shifts.
		\item For the observations, the deposited energy fluxes (derived using the thick-target model assumption) correlate with the 25-50\,keV photon fluxes averaged over the HXR sources. The linear correlation coefficient for their logarithmic values is very high (0.88, see Figure~\ref{figure4_photonflux_electronflux} and Table~\ref{table3_regression}). Strong deviations from this dependence may indicate on inapplicability of the thick-target model for some flare events.
		\item Despite differences in the slopes and low-energy cutoffs of the deposited energy electron spectra, RADYN models reveal linear dependence of the C\,II and Fe\,XXI peak Doppler shifts from the logarithm of the deposited energy flux, with high statistically-significant correlation coefficients (-0.73 for C\,II and 0.84 for Fe\,XXI).
		\item The only observational relation having a tendency to be statistically-significant is found for the strongest C\,II Doppler shifts and deposited energy flux. The empirical relation Eq.~3 (Table~\ref{table3_regression}) is the best candidate for the energy flux diagnostics from the UV spectroscopic data.
		\item The C\,II and Fe\,XXI line Doppler shifts derived for the studied flares do not correlate with each other in observations but do strongly correlate in RADYN models (the correlation coefficients are -0.13/-0.19 for observational results and -0.92 for the results of the models).
		\item The energy flux required for the transition from ``gentle'' to ``explosive'' evaporation regime is (2.2-10.1)$\cdot{}$10$^{9}$~erg~cm$^{-2}$~s$^{-1}$ from the RADYN models, and (2-8)$\cdot{}$10$^{9}$~erg~cm$^{-2}$~s$^{-1}$ from the observations. The observational estimate require verification on a larger statistics of events.
		\item There are qualitative discrepancies between the observations and RADYN models:
			\subitem a) The observed Fe\,XXI Doppler shifts are weaker than ones derived from the models. The maximum observed Fe\,XXI Doppler shifts reach 220\,km/s, while the models show the Doppler shifts of 400\,km/s and higher. However, notice that the synthesized Fe\,XXI emission is integrated over the whole loop, and it is likely that Fe\,XXI is unobservable during the most blueshifted phase due to low plasma emission measure at high ($>1\,MK$) temperatures at that moment.
			\subitem b) The observed mean C\,II Doppler shifts are always negative (corresponding to redshifts) during the studied flares even for the events with relatively weak deposited energy fluxes (of about 10$^{9}$~erg~cm$^{-2}$~s$^{-1}$). Contrary, the models predict that the C\,II Doppler shift should change from negative to positive with the decrease of the energy flux. However, for all events there are areas with blueshifted C\,II line during the impulsive phase of flares, which is in agreement with the multi-thread nature of solar flares.
	\end{enumerate}
	
	There are several assumptions made in this study. First, the IRIS raster scans cover only a part of the flare ribbons, and we assume that the distribution of chromospheric responses is the same in covered and uncovered parts of the ribbons. Second, we consider only certain descriptors of the UV line Doppler shifts but not their dynamical properties. Third, the synthetic spectral line profiles are calculated under assumption of statistical equilibrium for the C\,II line and ionization equilibrium optically-thin emission assumption for the Fe\,XXI line. Under these assumptions, our observational statistical study demonstrated for the first time how the Doppler shifts of UV lines during the chromospheric evaporation process depend on the deposited heat flux. Further joint X-ray and UV spectroscopic observations of flares as well as development of mode sophisticated data analysis techniques are needed for better understanding of the flare energy release and transport.

\acknowledgments

IRIS is a NASA small explorer mission developed and operated by LMSAL with mission operations executed at NASA Ames Research center and major contributions to downlink communications funded by ESA and the Norwegian Space Centre. We acknowledge the RHESSI team for the availability of the high-quality scientific data. The research leading to these results has received funding from the European Community’s Seventh Framework Programme (FP7/2007-2013) under grant agreement no. 606862 (F-CHROMA). We also acknowledge the Stanford Solar Observatories Group and NASA Ames Research Center for the possibility to use the computational resources. GSK is supported by an appointment to the NASA Postdoctoral Program at Goddard Space Flight Center, administered by USRA through a contract with NASA. The research was partially supported by the NASA Grants NNX14AB68G and NNX16AP05H, and NSF grant 1639683.

\bibliographystyle{aasjournal}

\bibliography{PaperStatisticalStudyOfChromorphericEvaporation}

\clearpage

\clearpage

\clearpage

\clearpage

\clearpage

\clearpage

\clearpage

\clearpage

\begin{sidewaystable}
	\centering
	\caption{Characteristics of the analyzed flares from spectroscopic analysis of UV and X-ray radiation. Parameters from the RHESSI X-ray data are obtained using thick-target spectral fitting and images reconstructed using ``vis\_cs'' algorithm. The spectral line Doppler shift characteristics are deduced from the IRIS UV data.}
	\label{table1_observations_viscs}
	\tiny
	\begin{tabular}{|c|c|c|c|c|c|c|c|}
		\hline
												&	SOL1				&	SOL2				&	SOL3				&	SOL4				&	SOL5				&	SOL6				&	SOL7	\\
												&	2014-02-13 01:32	&	2014-02-13 02:41	&	2014-03-29 17:35	&	2014-06-12 18:03	&	2014-06-13 00:30	&	2015-03-11 11:21	&	2015-08-27 04:48	\\
		\hline
		GOES class								&	M1.8				&	M1.0				&	X1.0				&	M1.3				&	C8.5				&	C5.8				&	M2.9	\\
		\hline
		RHESSI analysis times					&	01:34:35-01:36:05	&	02:47:35-02:48:35	&	17:44:55-17:45:55	&	18:07:30-18:08:30	&	00:33:30-00:34:30	&	11:26:00-11:28:00	&	05:36:20-05:37:20	\\
		\hline
		RHESSI detectors						&	1, 3, 4, 5, 7-9		&	1, 3, 4, 5, 7-9		&	2, 4, 5, 7-9		&	2, 4, 5, 7-9		&	1, 2, 4, 5, 7		&	1, 3, 5, 7-9		&	1, 5-9	\\
		\hline
		Fitting energy range [keV]				&	6-40				&	6-40				&	6-100				&	6-49				&	6-70				&	6-40				&	6-58	\\
		\hline
		$F_{\textrm{nonth}}$[$\textrm{erg}/\textrm{cm}^{2}\textrm{s}$]				&	(9.6$\pm{}$1.4)$\cdot$10$^9$	&	(3.9$\pm{}$1.1)$\cdot$10$^8$	&	(2.2$\pm{}$1.1)$\cdot$10$^{10}$	&	(9.0$\pm{}$2.6)$\cdot$10$^8$	&	(2.6$\pm{}$1.1)$\cdot$10$^9$	&	(3.1$\pm{}$0.8)$\cdot$10$^8$	&	(4.8$\pm{}$0.3)$\cdot$10$^{10}$	\\
		\hline
		$E_{c}$[keV]							&	19.4$\pm{}$0.1		&	18.8$\pm{}$0.3		&	17.0$\pm{}$0.3		&	19.0$\pm{}$0.4		&	17.2$\pm{}$0.1		&	22.2$\pm{}$0.9		&	19.8$\pm{}$0.1	\\
		\hline
		$\delta$								&	8.93$\pm{}$0.25		&	9.09$\pm{}$0.13		&	4.65$\pm{}$0.15		&	6.69$\pm{}$0.47		&	6.50$\pm{}$0.29		&	9.81$\pm{}$0.56		&	8.35$\pm{}$0.11	\\
		\hline
		C\,II mean Doppler shift [$\textrm{km/s}$]		&	-21$\pm{}$27		&	-14$\pm{}$17		&	-24$\pm{}$33		&	-32$\pm{}$30		&	-54$\pm{}$48		&	-10$\pm{}$13		&	-40$\pm{}$48	\\
		\hline
		C\,II mean Doppler shift time			&	01:36:18-01:36:56	&	02:47:34-02:48:12	&	17:45:36-17:46:41	&	18:08:21-18:08:39	&	00:34:10-00:34:28	&	11:27:04-11:28:09	&	05:37:25-05:37:46	\\
		\hline
		C\,II strongest redshifts [$\textrm{km/s}$]		&	-85$\pm{}$3			&	-40$\pm{}$3			&	-84$\pm{}$4			&	-90$\pm{}$5			&	-156$\pm{}$5		&	-31$\pm{}$4			&	-169$\pm{}$19	\\
		\hline
		C\,II strongest redshifts time			&	01:37:01-01:37:39	&	02:47:34-02:48:12	&	17:45:36-17:46:41	&	18:08:21-18:08:39	&	00:33:49-00:34:07	&	11:25:48-11:26:54	&	05:38:14-05:38:35	\\
		\hline
		Fe\,XXI mean Doppler shift [$\textrm{km/s}$]		&	43$\pm{}$27			&	17$\pm{}$18			&	42$\pm{}$34			&	142$\pm{}$42		&	27$\pm{}$24			&	42$\pm{}$25			&	53$\pm{}$47S	\\
		\hline
		Fe\,XXI mean Doppler shift time			&	01:36:18-01:36:56	&	02:48:17-02:48:55	&	17:45:36-17:46:41	&	18:07:59-18:08:18	&	00:33:49-00:34:07	&	11:25:48-11:26:54	&	05:37:25-05:37:46	\\
		\hline
		Fe\,XXI strongest blueshifts [$\textrm{km/s}$]	&	112$\pm{}$6			&	54$\pm{}$6			&	181$\pm{}$52		&	217$\pm{}$9			&	70$\pm{}$9			&	88$\pm{}$8			&	146$\pm{}$7		\\
		\hline
		Fe\,XXI strongest blueshifts time		&	01:36:18-01:36:56	&	02:48:17-02:48:55	&	17:44:21-17:45:27	&	18:08:21-18:08:39	&	00:33:49-00:34:07	&	11:27:04-11:28:09	&	05:37:25-05:37:46	\\
		\hline
	\end{tabular}
\end{sidewaystable}
\clearpage

\begin{table}
	\centering
	\caption{Characteristics of spectral lines calculated from the RADYN models.}
	\label{table2_modeling}
	\tiny
	\begin{tabular}{cccccc}
		\hline
		F-CHROMA ID				&	$F_{\textrm{nonth}}$[$\textrm{erg}/\textrm{cm}^{2}\textrm{s}$]	&	$E_{c}$[keV]	&	$\delta$	&	C\,II Doppler shift	&	Fe\,XXI Doppler shift	\\
								&								&					&				&	maximum, $\textrm{km/s}$		&	maximum, $\textrm{km/s}$	\\
		\hline
		d4\_3.0e10\_t20s\_15keV	&	1.5$\cdot$10$^9$			&	15				&	4			&	38.2				&	99.0	\\
		d7\_3.0e10\_t20s\_15keV	&	1.5$\cdot$10$^9$			&	15				&	7			&	44.7				&	71.0	\\
		d8\_3.0e10\_t20s\_15keV	&	1.5$\cdot$10$^9$			&	15				&	8			&	39.7				&	54.0	\\
		d7\_3.0e10\_t20s\_20keV	&	1.5$\cdot$10$^9$			&	20				&	7			&	17.2				&	---	\\
		d8\_3.0e10\_t20s\_20keV	&	1.5$\cdot$10$^9$			&	20				&	8			&	12.6				&	---	\\
		d4\_3.0e10\_t20s\_25keV	&	1.5$\cdot$10$^9$			&	25				&	4			&	8.4					&	---	\\
		d7\_1.0e11\_t20s\_15keV	&	5.0$\cdot$10$^9$			&	15				&	7			&	-4.6				&	400.0	\\
		d8\_1.0e11\_t20s\_15keV	&	5.0$\cdot$10$^9$			&	15				&	8			&	-11.4				&	422.0	\\
		d4\_1.0e11\_t20s\_20keV	&	5.0$\cdot$10$^9$			&	20				&	4			&	28.9 				&	148.0	\\
		d7\_1.0e11\_t20s\_20keV	&	5.0$\cdot$10$^9$			&	20				&	7			&	27.0				&	94.0	\\
		d8\_1.0e11\_t20s\_20keV	&	5.0$\cdot$10$^9$			&	20				&	8			&	39.7 				&	73.0	\\
		d8\_1.0e11\_t20s\_25keV	&	5.0$\cdot$10$^9$			&	25				&	8			&	17.4 				&	---	\\
		d8\_3.0e11\_t20s\_15keV	&	1.5$\cdot$10$^{10}$			&	15				&	8			&	-44.7				&	510.0	\\
		d4\_3.0e11\_t20s\_20keV	&	1.5$\cdot$10$^{10}$			&	20				&	4			&	8.3 				&	436.0	\\
		d8\_3.0e11\_t20s\_20keV	&	1.5$\cdot$10$^{10}$			&	20				&	8			&	-17.8				&	559.0	\\
		d6\_3.0e11\_t20s\_25keV	&	1.5$\cdot$10$^{10}$			&	25				&	6			&	11.4				&	242.0	\\
		d8\_3.0e11\_t20s\_25keV	&	1.5$\cdot$10$^{10}$			&	25				&	8			&	10.9				&	184.0	\\
		d4\_1.0e12\_t20s\_20keV	&	5.0$\cdot$10$^{10}$			&	20				&	4			&	-30.5				&	614.0	\\
		d5\_1.0e12\_t20s\_20keV	&	5.0$\cdot$10$^{10}$			&	20				&	5			&	-36.9				&	705.0	\\
		d5\_1.0e12\_t20s\_25keV	&	5.0$\cdot$10$^{10}$			&	25				&	5			&	-15.7				&	706.0	\\
		\hline
	\end{tabular}
\end{table}
\clearpage

\begin{sidewaystable}
	\centering
	\caption{Correlation coefficients and relationships for the observed and modeling parameters: non-thermal energy flux, $F_{\textrm{nonth}}$; 25-50\,keV photon flux, $F_{\textrm{ph}}$; C\,II mean Doppler shift, $v^{\textrm{C\,II}}_{\textrm{mean}}$; C\,II strongest redshifts, $v^{\textrm{C\,II}}_{5\% red}$; Fe\,XXI mean Doppler shift, $v^{\textrm{Fe\,XXI}}_{\textrm{mean}}$; Fe\,XXI strongest blueshifts, $v^{\textrm{Fe\,XXI}}_{\textrm{5\% blue}}$, presented in Figs~\ref{figure4_photonflux_electronflux}-\ref{figure7_fexxiblue_ciired} ($v$ in km/s, $F_{\textrm{ph}}$ in photons/cm$^{2}$arcsec$^{2}$s, $F_{\textrm{nonth}}$ in erg/cm$^{2}$s).\\
	$^{*}$p-value for a hypothesis test whose null hypothesis is an absence of association (Kendall's $\tau$ is zero). \\
	$^{+}$p-value for a hypothesis test whose null hypothesis is that the slope is zero.}
	\label{table3_regression}
	\tiny
	\begin{tabular}{|c|c|c|c|c|c|}
		\hline
		Pair of parameters	& Kendall's $\tau$	&	Kendall's $\tau$ p-value$^{*}$	&	Linear fit CC	&	Linear fit p-value$^{+}$	&	Linear fit equation	\\
		\hline
		\multicolumn{6}{|c|}{Observational parameters}	\\
		\hline
		$F_{\textrm{ph}}$ and $F_{\textrm{nonth}}$	&	0.90	&	0.0043	&	0.88	&	0.0083	&	$log_{10}F_{\textrm{nonth}} = (11.6\pm{}0.5) + (0.87\pm{}0.20)\cdot{}log_{10}F_{\textrm{ph}}$ (2)	\\
		$F_{\textrm{nonth}}$ and $v^{\textrm{C\,II}}_{\textrm{mean}}$	&	-0.52	&	0.10	&	-0.42	&	0.34	&	~---	\\
		$F_{\textrm{nonth}}$ and $v^{\textrm{C\,II}}_{5\% red}$	&	-0.52	&	0.10	&	-0.68	&	0.093	&	$v^{\textrm{C\,II}}_{\textrm{5\% red}} = (302\pm{}195) - (42\pm{}21)\cdot{}log_{10}F_{\textrm{nonth}}$ (3)	\\
		$F_{\textrm{nonth}}$ and $v^{\textrm{Fe\,XXI}}_{\textrm{mean}}$	&	0.29	&	0.36	&	-0.08	&	0.87	&	~---	\\
		$F_{\textrm{nonth}}$ and $v^{\textrm{Fe\,XXI}}_{\textrm{5\% blue}}$	&	0.33	&	0.29	&	0.39	&	0.38	&	$v^{\textrm{Fe\,XXI}}_{\textrm{5\% blue}} = (-138\pm{}274) + (28\pm{}29)\cdot{}log_{10}F_{\textrm{nonth}}$ (4)	\\
		$v^{\textrm{C\,II}}_{\textrm{mean}}$ and $v^{\textrm{Fe\,XXI}}_{\textrm{mean}}$	&	-0.20	&	0.54	&	-0.13	&	0.78	&	~---	\\
		$v^{\textrm{C\,II}}_{5\% red}$ and $v^{\textrm{Fe\,XXI}}_{\textrm{5\% blue}}$	&	-0.24	&	0.45	&	-0.19	&	0.69	&	~---	\\
		\hline
		\multicolumn{6}{|c|}{Modeling parameters}	\\
		\hline
		$F_{\textrm{nonth}}$ and $v^{\textrm{C\,II}}$	&	-0.58	&	0.0012	&	-0.73	&	2.4$\cdot{}10^{-4}$	&	$v^{\textrm{C\,II}} = (359.0\pm{}77.0) - (35.9\pm{}7.9)\cdot{}log_{10}F_{\textrm{nonth}}$ (5)	\\
		$F_{\textrm{nonth}}$ and $v^{\textrm{Fe\,XXI}}$	&	0.77	&	0.00012	&	0.84	&	4.6$\cdot{}10^{-5}$	&	E$v^{\textrm{Fe\,XXI}} = (3500.0\pm{}662.0) + (386.0\pm{}66.0)\cdot{}log_{10}F_{\textrm{nonth}}$ (6)	\\
		$v^{\textrm{C\,II}}$ and $v^{\textrm{Fe\,XXI}}$	&	-0.78	&	2.8$\cdot{}10^{-5}$	&	-0.92	&	4.3$\cdot{}10^{-7}$	&	$v^{\textrm{Fe\,XXI}} = (373.0\pm{}24.5) - (7.45\pm{}0.84)\cdot{}v^{\textrm{C\,II}}$ (7)	\\
		\hline
	\end{tabular}
\end{sidewaystable}
\clearpage

\end{document}